# Topological triply-degenerate point with double Fermi arcs

*Yihao Yang[1,2,#], Hong-xiang Sun[3,#], Jian-ping Xia[3], Haoran Xue[1,2], Zhen Gao[1,2], Yong Ge[3], Ding Jia[3], Yidong Chong[1,2,*], and Baile Zhang[1,2,*]*

[1]Division of Physics and Applied Physics, School of Physical and Mathematical Sciences, Nanyang Technological University, 21 Nanyang Link, Singapore 637371, Singapore.

[2]Centre for Disruptive Photonic Technologies, The Photonics Institute, Nanyang Technological University, 50 Nanyang Avenue, Singapore 639798, Singapore.

[3]Faculty of Science, Jiangsu University, Zhenjiang 212013, China.

#These authors contributed equally to this work.

*Correspondence to: (Y. C.) yidong@ntu.edu.sg; (B. Z.) blzhang@ntu.edu.sg.

## ABSTRACT

**Unconventional chiral particles have recently been predicted to appear in certain three dimensional (3D) crystal structures containing three- or more-fold linear band degeneracy points (BDPs). These BDPs carry topological charges, but are distinct from the standard twofold Weyl points or fourfold Dirac points, and cannot be described in terms of an emergent relativistic field theory. Here, we report on the experimental observation of a topological threefold BDP in a 3D phononic crystal. Using direct acoustic field mapping, we demonstrate the existence of the threefold BDP in the bulk bandstructure, as well as doubled Fermi arcs of surface states consistent with a topological charge of 2. Another novel BDP, similar to a Dirac point but carrying nonzero topological charge, is connected to the threefold BDP via the doubled Fermi arcs. These findings pave the way to using these unconventional particles for exploring new emergent physical phenomena.**

It is a remarkable fact that particle-like excitations described by relativistic field theories can arise in non-relativistic periodic media whose bandstructures contain linear band degeneracy points (BDPs). For example, two-dimensional (2D) hexagonal crystals like graphene can exhibit stable twofold linear BDPs, called Dirac points, whose excitations act as 2D massless spin-1/2 Dirac particles[1]. In three dimensions (3D), the analogous twofold linear BDPs are called Weyl points, and their excitations are massless particles of a type hypothesized but so far not known to exist in high-energy particle physics, i.e., spin-1/2 Weyl particles[2-10]. Weyl points carry topological charges (Chern numbers) of either +1 or -1[2-10], corresponding to the chirality (left- or right-handed) of the Weyl particles. As a result, each pair of oppositely charged Weyl points, when projected on a surface, is joined by an unclosed arc, called a Fermi arc, of topologically-protected surface states. A 3D Dirac point, on the other hand, is a fourfold BDP that can be viewed as two overlapping Weyl points with opposite topological charges; it carries no net topological charge and thus lacks topological surface states[4,11]. Weyl and Dirac particles obey the Lorentz group of free-space relativistic transformations, and can be regarded as emergent relativistic particles with the group velocity of the linear BDP playing the role of the speed of light.

Recently, it has been predicted that certain bandstructures can have symmetry-stabilized linear BDPs that host a family of unconventional chiral quasiparticles beyond Weyl and Dirac particles[12-15]. These "unconventional BDPs" occur in nonsymmorphic crystal structures and carry nonzero topological charge[12]. One such candidate is a threefold linear BDP (henceforth called the "triple point") carrying topological charge 2. This is qualitatively distinct from twofold BDPs generated by merging a pair of Weyl points of the same charge, which have quadratic or higher-order dispersion relations[16-18]. The excitations of the triple point behave as spin-1 quasiparticles, unlike Weyl and Dirac particles which are spin-1/2. Because the triple point has a topological charge of 2, there are two topologically-protected Fermi arcs emanating simultaneously from its projection onto a 2D surface. This topological charge distinguishes the present triple point from recently-observed threefold BDPs in MoP and WC crystals[19,20], which do not carry topological charge, similar to 3D Dirac points. A threefold degeneracy has also been observed in the phonon spectrum of FeSi[21], but at zero frequency, which is not a BDP in the sense considered in Ref. [12]. To our knowledge, the double Fermi arcs predicted to occur for topologically nontrivial threefold BDPs[12] has never been observed.

We have designed and implemented an acoustic metamaterial – a classical 3D phononic crystal with a nonsymmorphic structure – featuring a charge-2 triple point. By direct field-mapping, we demonstrate experimentally that the three affected acoustic bands have the desired dispersion characteristics at the center of the Brillouin zone (Γ). Another interesting bandstructure feature observed in these measurements is a charge-2 fourfold BDP (henceforth called the "quadruple point") existing at the corner of the Brillouin zone. It is equivalent to a pair of Weyl points with the *same* topological charge, unlike standard 3D Dirac points which are equivalent to a pair of oppositely-charged Weyl points (thus carrying zero topological charge). Moreover, we are able to verify the topological charges of both BDPs by imaging, for the first time, the hallmark "double Fermi arcs" that extend from the triple point to the quadruple point. These double Fermi arcs form double helicoids over a broad frequency range (relative bandwidth >25%), and are thus experimentally very robust. We show that the Fermi arc surface states exhibit the distinctive phenomenon of topologically protected negative refraction when propagating across corners of the sample.

The sample of 3D phononic crystal, shown in Figs. 1A-B, consists of 20×20×10 unit cells in a cubic lattice structure, with the noncentrosymmetric space group $P2_13$ (No. 198). The lattice constant is $a$=20 mm.

As shown in Fig.1C, each unit cell contains four junctions, at ($u$, $u$, $u$), ($0.5a+u$, $0.5a-u$, $-u$), ($-u$, $0.5a+u$, $0.5a-u$), and ($0.5a-u$, $-u$, $0.5a+u$), where $u=0.1a$. Each junction connecting to 6 neighbouring junctions via solid rods of radius $r=0.14a$; the rest of the volume is filled with air. The resulting crystal structure has three twofold screw symmetries, along each <100> axis, and a $C_3$ rotational symmetry along the <111> axes. The sample is fabricated from photosensitive resin via 3D printing, and is self-supporting.

The bandstructure for acoustic (sound) waves in this lattice was calculated numerically, and the results are shown in Figs. 1D-F. There exists a triple point at the Γ point in the center of the Brillouin zone, and a quadruple point at the R point in the corner of the Brillouin zone. The triple point can be described by a three-band effective Hamiltonian $H_3(k) \sim k \cdot L$, where $k$ is the wavevector and $L$ is the spin-1 matrix representation of the rotation generator[12-15], and the Chern numbers of the bands are +2, 0, and -2, respectively. As for the quadruple point, it can be treated as a direct sum of two identical Weyl points, via the four-band effective Hamiltonian $H_4(k) \sim \begin{pmatrix} k.\sigma & 0 \\ 0 & k.\sigma \end{pmatrix}$, where $\sigma$ denotes the Pauli matrices[13-15]; the bands have Chern numbers of +2 and -2, respectively. All Chern numbers were calculated numerically from the acoustic Bloch wavefunctions using the Wilson loop method[22] (see Supplementary Information for details). The two BDPs have opposite topological charges ±2 that sum to zero, consistent with the Nielsen-Ninomiya theorem[23]. Note that although the BDPs occur at different frequencies, there are no trivial bands occupying the frequency range between them. Therefore, this phononic crystal provides an extremely "clean" platform for studying the physical effects of the BDPs.

We determine the bulk bandstructure experimentally by threading acoustic sources and probes through the air holes of the phononic structure, and measuring the pressure field distribution along a horizontal plane in the middle of the sample, with a source fixed 4 cm below the plane (Fig. 2A). After performing a 2D Fourier transform, we obtain the projected band structure (projected from 3D into the 2D plane) shown in Fig. 2B. The results reveal, as expected, (i) a nodal point (the triple point) at 13.4 kHz projected onto $\bar{\Gamma}$ in the 2D Brillouin zone, and (ii) another nodal point (the quadruple point) at 11.2 kHz projected onto $\bar{M}$ . This is in excellent quantitative agreement with the numerically-calculated projected bandstructure shown in Fig. 2C.

Next, we use surface measurements to verify the topological charges of the BDPs. The setup for the surface measurements is shown in Fig. 3A. The phononic crystal is placed on an acrylic board that behaves as a hard boundary for sound. A source is placed at the center of the bottom surface, exciting acoustic surface waves. We then insert a probe into the sample from the top, and measure the acoustic field distribution just above the surface. After applying a 2D Fourier transform, we obtain the surface spectrum shown in Figs. 3B, D, and F, which agree very well with the numerically-calculated surface dispersion (Figs. 3C, E, and G). It is evident from the measurement results that there exist two Fermi arcs, consistent with the predicted topological charges of the BDPs. Over frequencies ranging from 10 kHz to 13 kHz, the dispersion surface forms a double helicoid: two surface sheets winding around the triple and quadruple points. Interestingly, the isofrequency curve forms a single noncontractible loop (Figs. 3D and F) wrapping around the torus of Brillouin zone, which is qualitatively different from those of trivial surface states (whose isofrequency curves are closed loops which can deform into an even number of noncontractible loops), and also different from the non-trivial surface states of conventional Weyl points (whose dispersion surfaces are single helicoids, and isofrequency curves are open arcs)[13] .

Finally, we make use of the robustness of the Fermi arc surface states by demonstrating negative surface refraction across two surfaces of the sample. As shown in Figs. 4A-B, the adjacent surfaces are covered with acrylic boards (see the schematic illustration in Supplementary Information), a source is fixed on one surface, and the field distribution is measured along both surfaces. When the surface states impinge the edge between the two surfaces, no reflection occurs, due to the lack of a reflection channel. The refracted wave on the other surface lies on the same side of the normal as the incident wave, a phenomenon referred to as "negative refraction" for topological surface waves[24]. This can be explained from the isofrequency contour of the surface states. It can be seen from Figs. 1A-C that the unit cells on the two surfaces are identical apart from a 90° rotation. Thus, the isofrequency contours on the two surfaces rotated by 90° relative to each other, as shown in Fig. 4C, resulting in negative refraction. Note that this 90° rotation of isofrequency contours is applicable to all six surfaces of the sample; hence, the negative refraction phenomenon can occur at all twelve edges.

Our work thus experimentally demonstrates two novel types of topologically nontrivial linear BDPs: a charge-2 triple point and a charge-2 quadruple point. The former is notable because it is an unconventional BDP that lies outside the pre-existing Weyl/Dirac types and corresponds to spin-1 chiral particles beyond the standard model. As for the latter, it can be regarded as a merger of two Weyl points but is also notable because, unlike previously-studied 3D Dirac points, the Weyl points have the same topological charge. We have shown that the BDPs are connected by double Fermi arcs, which can be observed over a broad range of frequencies and is responsible for the distinctive phenomenon of negative surface refraction. As the triple point lies at the center of the Brillouin zone, it may be interpretable in terms of a three-dimensional zero-index metamaterial[25] – but which, unlike previously-studied zero-index metamaterials based on fine-tuned accidental degeneracies, is symmetry-stabilized and thus extremely robust[26]; this may have interesting implications for three-dimensional waveguiding. Due to the ease with which phononic crystals can be fabricated and analysed, it will be interesting to use them to look for other types of unconventional BDPs, which are stabilized by different choices of crystal symmetry and can carry topological charges higher than ±2, and thus more intricate surface state dispersions.

## Acknowledgments

We thank H. S. Tan at Nanyang Technological University for helpful discussions. This work was sponsored by Singapore Ministry of Education under Grants No. MOE2015-T2-1-070, MOE2015-T2-2-008, MOE2016-T3-1-006 and Tier 1 RG174/16 (S). H.S. acknowledges the support of the National Natural Science Foundation of China (11774137).

## Authors Contributions

All authors contributed extensively to this work. Y.Y. designed the phononic crystal and performed the simulations. Y.Y., H.X., Z.G. and H.S. fabricated the sample and designed the experiments. H.S., J.X., Y.G. and D.J. performed measurements. Y.Y., Y.C. and B.Z. analyzed data and wrote the paper. Y.C. and B.Z. supervised the project.

## Competing Financial Interests

The authors declare no competing financial interests.

## Data availability

The data that support the findings of this study are available from the corresponding author upon reasonable request.

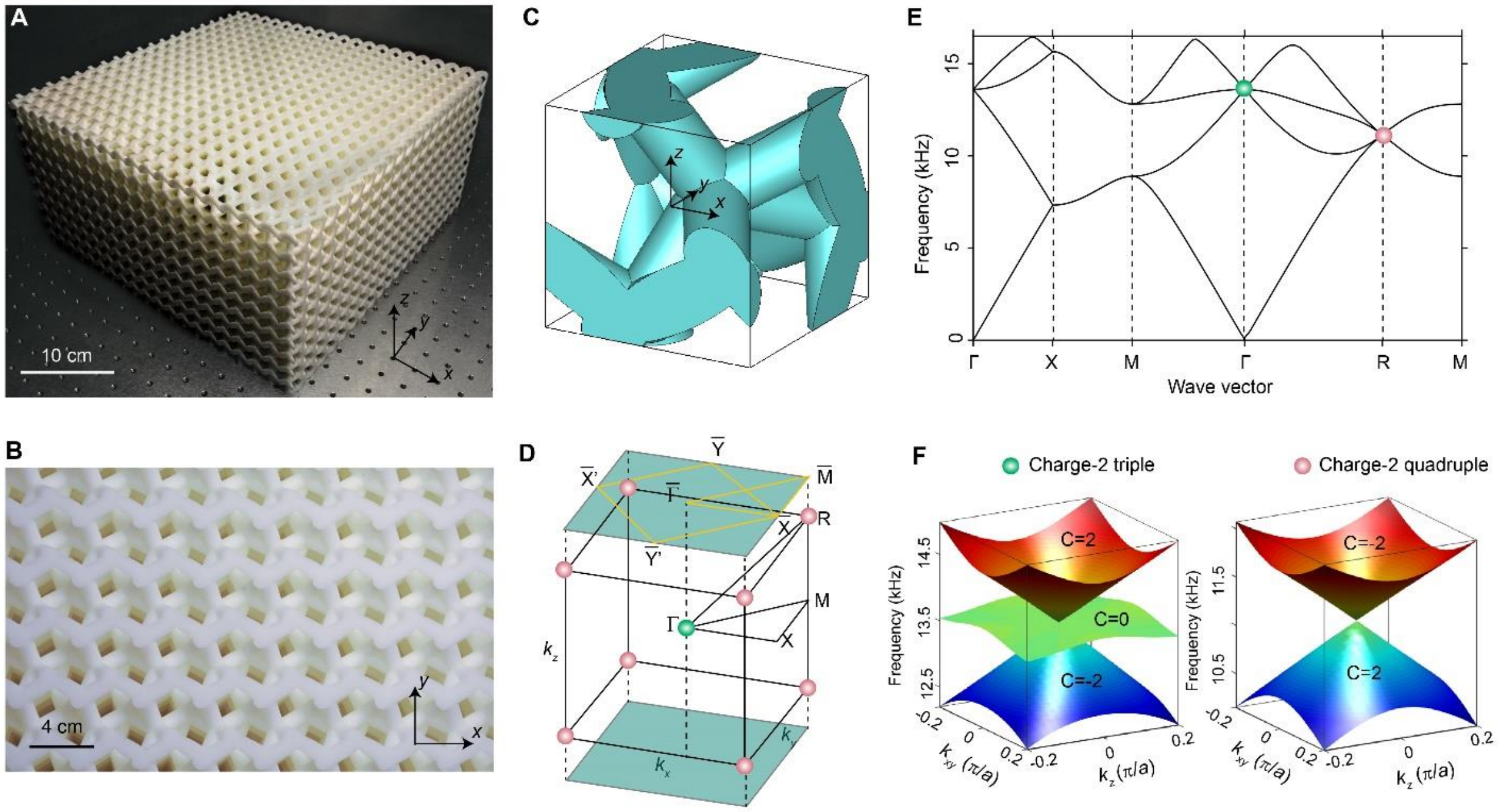

**Figure 1. 3D phononic crystal with a charge-2 triple point and a charge-2 quadruple point.** (**A**-**B**) Photographs of the fabricated sample. (**C**) Unit cell of the phononic crystal. (**D**-**E**) Brillouin zone and band structure of the phononic crystal, with red and green points indicating the triple point and quadruple point, respectively. (**F**) 2D band structures in the vicinity of the triple point and quadruple points. The Chern number of each band is labeled.

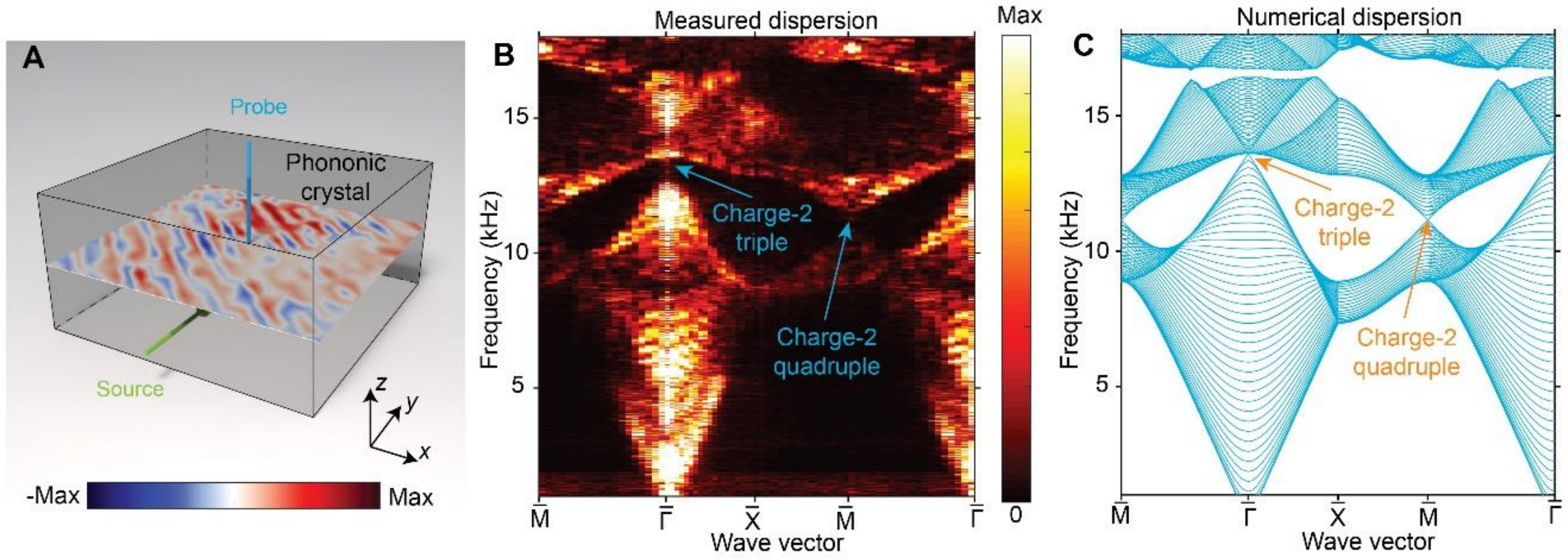

**Figure 2. Experimental observation of a charge-2 triple point and a charge-2 quadruple point.** (A) Experimental setup. The field distributions are measured along the middle plane of the sample. (**B**) Measured projected band spectrum. The colour bar indicates the energy intensity. (**C**) Numerically-calculated projected band spectrum.

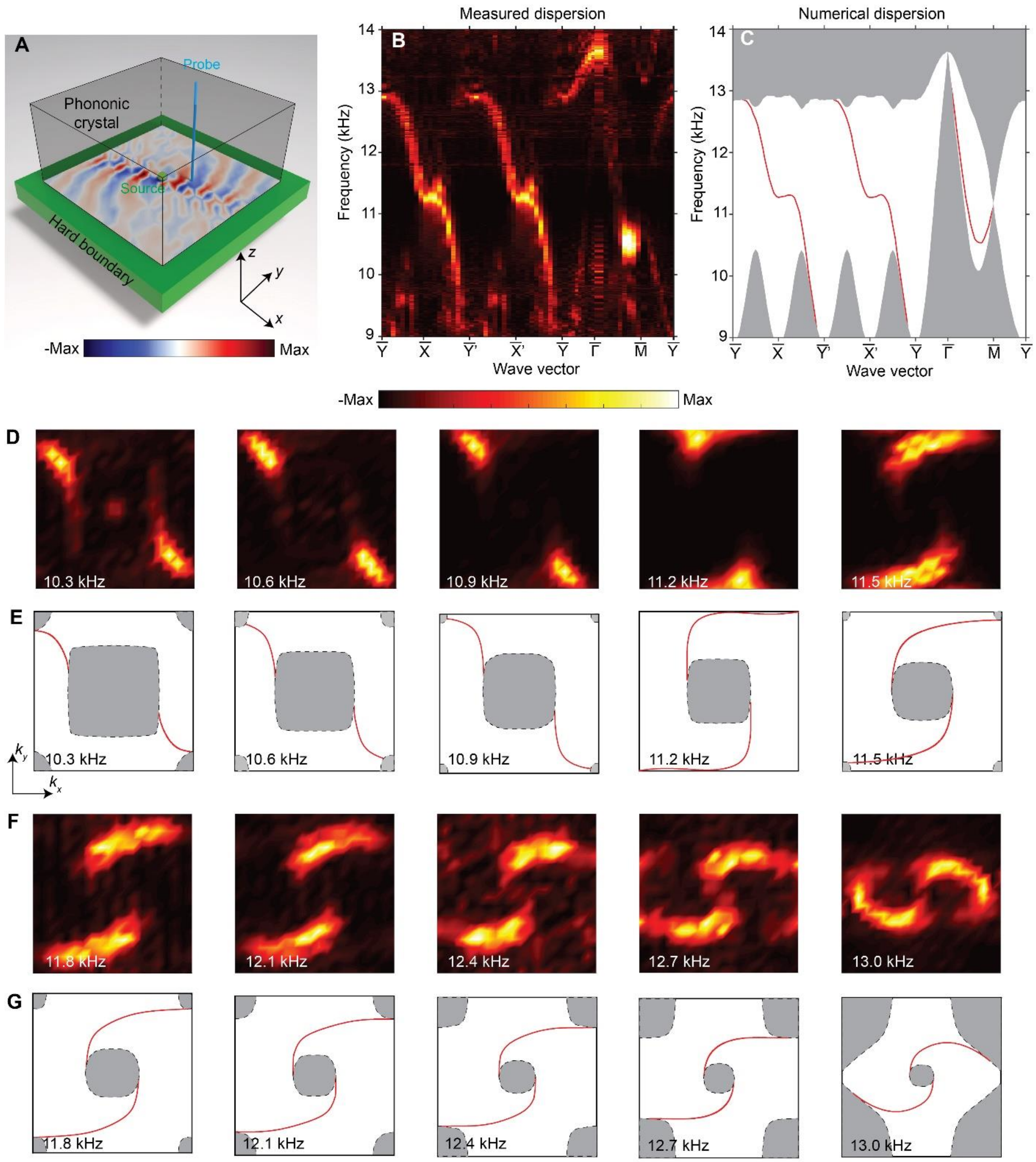

**Figure 3. Experimental observation of the double-helicoid topological surface states.** (**A**) Experimental setup. The field distributions are measured along the bottom surface of the sample. The sample sits on an acrylic board that acts as a hard boundary for sound. (**B**) Measured surface dispersion of the phononic crystal along the high-symmetry lines. (**C**) Numerical surface dispersion of the phononic crystal along the high-symmetry lines. (**D** and **F**) Measured surface isofrequency contours from 10.3 kHz to 13.0 kHz. (**E** and **G**) Numerically-calculated surface isofrequency

contours from 10.3 kHz to 13.0 kHz. The plotted range for each panel is $[-\pi/a, \pi/a]^2$. The colour bar below (**B**) shows the energy intensity.

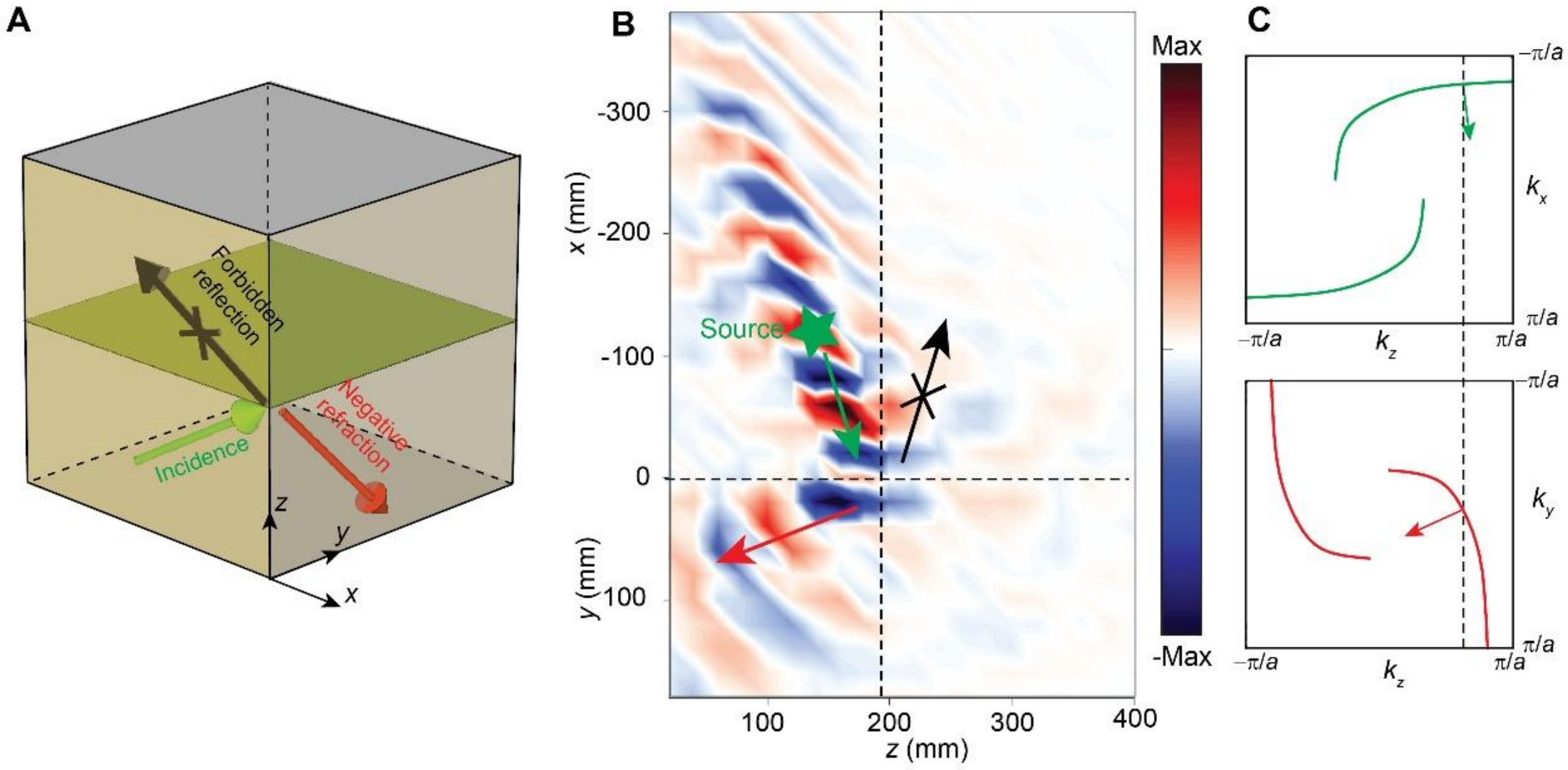

**Figure 4. Negative refraction of topological surface states.** (**A**) Schematic of the negative refraction experiment. The green, black, and red arrows represent incidence, reflection, and refraction respectively. The green plane denotes the normal. (**B**) Measured surface acoustic waves on the *y*=0 mm and *x*=0 mm planes at 11.8 kHz. The color bar measures the sound pressure. (**C**) Isofrequency contours for the surface states on the *y*=0 and *x*=0 planes, respectively.